\documentstyle[aps,prb,twocolumn,epsf,rotate]{revtex}

\begin{document}

\twocolumn[\hsize\textwidth\columnwidth\hsize\csname
@twocolumnfalse\endcsname

\title{Theory of Ferromagnetism in Diluted Magnetic Semiconductor Quantum Wells}

\draft

\author{ Byounghak Lee$^{1}$, T. Jungwirth$^{1,2}$, and A.H. MacDonald$^{1}$}
\address{$^{1}$Department of Physics,
Indiana University, Bloomington, Indiana 47405}
\address{$^{2}$Institute of Physics ASCR,
Cukrovarnick\'a 10, 162 00 Praha 6, Czech Republic}
\date{\today}
\maketitle

\begin{abstract}

We present a mean field theory of ferromagnetism in diluted magnetic
semiconductor  quantum wells.  When subband mixing due to
exchange interactions between quantum well free carriers and magnetic
impurities is neglected, analytic result can be obtained for
the dependence of the critical
temperature and the spontaneous magnetization on the distribution
of magnetic impurities and the quantum well width. 
The validity of this approximate theory has been tested
by comparing its predictions with those from numerical
self-consistent field calculations. Interactions among free carriers,
accounted for using the local-spin-density approximation, 
substantially  enhance the critical temperature. We demonstrate
that an external bias potential
can tune the critical temperature through a wide range.

\end{abstract}

\pacs{}


\vskip2pc]

The discovery of carrier-mediated
ferromagnetism\cite{story86,ohno92,ohno96,review}
in diluted magnetic semiconductors (DMS) has opened a broad and relatively
unexplored frontier for both basic and applied research.
The interplay between collective magnetic properties and
semiconductor transport properties in these systems presents a
rich phenomenology and offers the prospect of new functionality
in electronic devices.  The possibilities for manipulating this
interplay are especially varied when the free carrier system is
quasi-two-dimensional\cite{2dpolish,2dawschalom,2dsmorchkova} because of the dependence
of system properties on the subband wavefunction.  In this article
we address the dependence of the ferromagnetic critical temperature
$T_c$ of two-dimensional DMS ferromagnets on
the spatial distribution of magnetic ions, the subband
wavefunction of the free carrier system, and on their interplay.
Our analysis is based on a formulation of the mean-field theory
for carrier-induced ferromagnetism\cite{dietl} in a DMS which was
developed in earlier work\cite{usprb} and is intended to be useful
for any spatially inhomogeneous system.  We find that
$T_c$ is quite sensitive to the
relative distributions of electrons and magnetic ions and can
be altered {\it in situ} by applying a gate voltage.

For definiteness we address the case of a $\langle 001\rangle$ growth direction
2D hole gas in the Ga$_{1-x}$Mn$_{x}$As DMS system. 
Much of our analysis, however,
applies equally well to other interfaces and to other
DMS systems with cubic host semiconductors
and, with one important caveat which we mention below, also to the case
of a 2D electron gas mediated ferromagnetism.  Our theory is based
on an envelope function description of the valence band electrons
and a spin representation of their kinetic-exchange
interaction\cite{dmsreviews} with
d-electrons\cite{commentonmodel} on the $S=5/2$ Mn$^{++}$ ions:
\begin{equation}
{\cal H} = {\cal H}_m + {\cal H}_f + J_{pd} \sum_{i,I} {\vec S_I}
\cdot {\vec s}_i \delta({\vec r}_i - {\vec R}_I),
\label{coupling}
\end{equation}
where $i$ labels a free carrier, $I$ labels a magnetic
ion and the exchange interaction energy $J_{pd}\approx 0.15$~eV~nm$^3$.  
In Eq.(~\ref{coupling}) ${\cal H}_m$ is the Hamiltonian of the
magnetic ions, ${\cal H}_f$ is the four-band Luttinger Hamiltonian
for free carriers in the valence band,  $\vec S$ is the magnetic ion  spin and
$\vec s$ is the electron-spin operator projected onto the $j=3/2$
valence band manifold of the Luttinger Hamiltonian.
We assume here that the 2D free carrier density is sufficiently low that
only the lowest energy heavy-hole subband is occupied, and that the
quantum well  of interest is sufficiently narrow
that mixing between light-hole and heavy-hole bands can be
neglected.  These simplifications lead\cite{valencebands} to
a single parabolic band with the in-plane effective mass 
$m^*_{\parallel} \approx
0.11m_0$ and the out-of-plane mass $m^*_{z} \approx
0.38m_0$. The two spin-states in this band have definite $j_z=\pm 3/2$ and
definite $s_z = \pm 1/2$; the projection of the transverse spin components
onto this band is zero so that the kinetic exchange interaction
has an Ising character.\cite{caveatanisotropy}

Our  mean-field theory\cite{usprb} is 
derived in a spin-density-functional
theory framework and 
leads to a set of physically transparent coupled equations.
In the absence of external fields, the mean polarization of a
magnetic ion is given by
\begin{equation}
\langle m \rangle_I = S B_{S}\big(J_{pd} S \big[n_{\uparrow}(\vec R_I)
-n_{\downarrow}(\vec R_I)\big]/2 k_B T\big);
\label{mI}
\end{equation}
where $B_S(x)$ is the Brillouin function,
\begin{eqnarray}
B_S(x) & = & {2S+1 \over 2S} \coth({2S+1 \over 2S} x)
        - {1 \over 2S}\coth({ 1\over2S}x) \nonumber \\
& \approx & {S+1 \over 3S}x -{(S+1)(2S^2 +2S+1) \over 90S^3}x^3, \: x \ll
1\, .
\end{eqnarray}
The electron spin-densities $n_{\sigma}(\vec r)$ are 
determined by solving the Schr\"{o}dinger equation
for electrons which experience a spin-dependent potential
due to kinetic exchange with the polarized magnetic ions:
\begin{eqnarray}
& &\bigg[ - \frac{ 1}{2 m^*_{\parallel}}\big(\frac{\partial^2}{\partial x^2}
+\frac{\partial^2}{\partial y^2}\big)-
\frac{ 1}{2 m^*_{z}}\frac{\partial^2}{\partial z^2}
+ v_{es}(\vec r)  \nonumber \\ 
& &+v_{xc,\sigma}(\vec r)
 - \frac{\sigma}{2} h_{pd}(\vec r) \bigg] \psi_{k,\sigma}(\vec r)
 = \epsilon_{k,\sigma} \psi_{k,\sigma}(\vec r)\; ;
\label{kseqs}
\end{eqnarray}
\begin{equation}
n_{\sigma}(\vec r) = \sum_k f(\epsilon_{k,\sigma}) |\psi_{k,\sigma}(\vec
r)|^2\; .
\label{densities}
\end{equation}
In these equations $v_{es}(\vec r)$ is the electrostatic potential,
including band offset and ionized impurity contributions, 
$v_{xc,\sigma}(\vec r)$ is
an exchange-correlation potential on which we comment further
below, and $f(\epsilon)$ is the Fermi distribution function.  
The spin-dependent kinetic-exchange potential,
\begin{equation}
h_{pd}(\vec r) = J_{pd} \sum_I \delta (\vec r - \vec R_I) \langle m
\rangle_I
\label{exchangefield}
\end{equation}
is non-zero only in the ferromagnetic state.  In the following we
assume that the magnetic ions are randomly distributed
and dense on a scale set by the free carrier Fermi wavevector in the
$\hat x-\hat y$ plane, and that their density in the growth
direction, $c(z)$, can be precisely controlled.\cite{2dsmorchkova}
This allows us to take a continuum limit where the magnetic ion
polarization and the kinetic-exchange potential depends 
only on the growth direction coordinate
and
\begin{equation}
h_{pd}(z) = J_{pd} \  c(z)\ \langle m \rangle(z).
\label{continuum}
\end{equation}
In the following we discuss how the ferromagnetic transition
temperature $T_c$ depends on $c(z)$.

The $T_c$ calculation is greatly simplified when the spin-dependence of the
exchange correlation potential is neglected and
the quantum well width $w$ is small enough to make subband mixing due to kinetic
exchange interactions negligible, i.e., when $J c(z) \ll \hbar^2/m^{\ast}_zw^2$.
We see later that the theory which results from these 
approximations is normally
quite accurate.  With subband mixing neglected, the only effect
of the mean-field kinetic-exchange interaction on the 2D
carriers is to produce a rigid spin-splitting of the 2D bands,
given by the first-order perturbation theory expression:
\begin{equation}
\epsilon_{\downarrow}-\epsilon_{\uparrow} =
J_{pd} \int dz^{\prime} \langle m \rangle(z^{\prime}) c(z^{\prime}) |\psi(z^{\prime})|^2
\end{equation}
where $\psi(z)$ is the growth direction envelope function of the
lowest subband.  For the case of a quantum well with infinite
barriers, $\psi(z) = \sqrt{2/w} \cos({\pi z / w})$.
The electron spin-polarization can then be obtained by summing over occupied
free-carrier states.  It follows that, as long as both spin-$\uparrow$
and spin-$\downarrow$ bands are partly occupied,
\begin{equation}
n_{\uparrow}(z)-n_{\downarrow}(z)={m^{\ast}_{\parallel} \over 2 \pi \hbar^2} |\psi(z)|^2
J_{pd} \int dz^{\prime} \langle m \rangle(z^{\prime}) c(z^{\prime}) |\psi(z^{\prime})|^2 \:.
\label{spinpolarization}
\end{equation}
Eq.~(\ref{spinpolarization}) can be used to eliminate the free-carrier
spin-polarization from Eq.~(\ref{mI}) and to obtain a self-consistent equation for
the function $\langle m \rangle(z)$. The equation depends only on
the kinetic-exchange coupling constant $J_{pd}$, the free-carrier in-plane
effective mass $m^*_{\parallel}$, the subband wavefunction $\psi(z)$, and the
magnetic ion distribution function $c(z)$.  It is important to
realize that the entire system behaves collectively.  It has a single
critical temperature which depends on the function $c(z)$, rather than a z-dependent
$T_c$ like that which appears in a naive adaptations of the bulk
RKKY theory for $T_c$ to quasi-2D systems.

\begin{figure}[h]
\hspace*{-.1cm}
\epsfxsize=3.6in
{\centerline{\epsffile{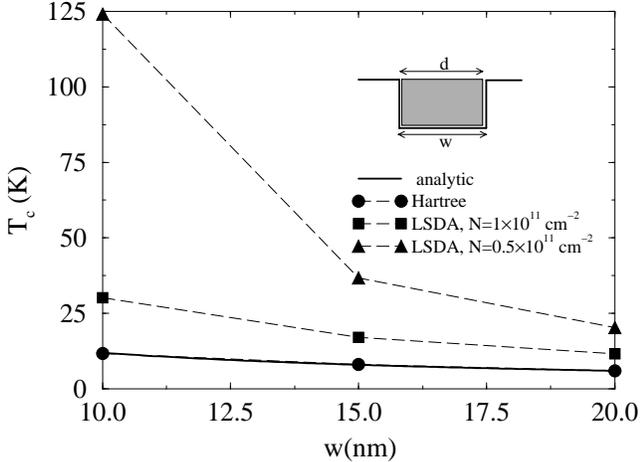}}}
\caption{Ferromagnetic critical temperature of a quasi-2D DMS as a 
function of quantum well width $w$  
for fixed 3D magnetic ion density
$c=1 {\rm nm}^{-3}$; inset: schematic diagram of the quantum well with
the shaded portion of width $d$ doped with magnetic ions. In this
figure $d=w$.
The non-interacting free carrier results were
calculated both by solving the full self-consistent equations (circles)
and by using the approximate $T_c$ expression derived in the text (solid line),
confirming the accuracy of the latter approach. Squares and triangles represent
the full numerical LSDA calculations for 2D carrier densities $N=1\times10^{11}$ cm$^{2}$
and $N=0.5\times10^{11}$ cm$^{2}$, respectively.}
\label{fig:one}
\end{figure}

An explicit expression for the critical temperature can be
obtained by linearizing the self-consistent equation (\ref{mI}).  
Expanding the Brillouin function
to the first order in its argument gives the homogeneous linear
integral equation
\begin{equation}
\langle m \rangle(z) = \frac{1}{k_B T} \int d z' K(z,z') \langle m
\rangle(z')
\label{tcequation}
\end{equation}
where the kernel
\begin{equation}
K(z,z') = {S (S+1) J_{pd}^2 \over 12} {m^{\ast}_{\parallel} \over \pi \hbar^2}
  |\psi(z)|^2  c(z^{\prime}) |\psi(z^{\prime})|^2.
\end{equation}
Since the linearized equation is satisfied only at the critical
temperature, $k_B T_c$ is equal to the largest eigenvalue of
$K(z,z')$, the eigenvalue corresponding to eigenfunction $f(z) \propto
|\psi(z)|^2$.  It follows that
\begin{equation}
T_c = {S (S+1) \over 12} {J_{pd}^2 \over k_B} {m^{\ast}_{\parallel} \over \pi \hbar^2}
\int dz |\psi(z)|^4 c(z) \:.
\label{tc}
\end{equation}
The critical temperature is proportional to $J_{pd}^2$ and $m^*_{\parallel}$, 
is independent of 2D carrier density, and
for a system with uniformly distributed magnetic impurities is
inversely proportional to the quantum well width, as illustrated 
by the solid line in 
Fig.~\ref{fig:one}.  The accuracy of Eq.~(\ref{tc}) has been tested by 
numerically solving the set of self-consistent Eqs.~(\ref{mI})-(\ref{continuum})
together with the Poisson equation for $v_{es}(z)$ and the 
local-spin-density-approximation (LSDA) formula\cite{vosko} for $v_{xc,\sigma}(z)$.
If  exchange and correlation effects are neglected ($v_{xc,\sigma}(z)=0$),
the self-consistent numerical results represented by circles in Fig.~\ref{fig:one}
are identical to those obtained from Eq.~(\ref{tc}) for quantum well widths $w=10$,
15, and 20~nm. Interactions among the 2D carriers substantially enhance the 
critical temperature and
also cause a strong dependence of $T_c$ on the 2D carrier density $N$, as illustrated
by squares ($N=1\times10^{11}$ cm$^{2}$) and triangles ($N=0.5\times10^{11}$ cm$^{2}$)
in  Fig.~\ref{fig:one}. Note that unlike the 3D case, where $T_c$ is proportional
to  Fermi wavevector in the non-interacting 
limit\cite{dietl,usprb} and is an increasing function
of density even if the interactions are included,\cite{usprb} 
the critical temperature for all three quantum wells of Fig.~\ref{fig:one}
is larger at  lower density, $N=0.5\times10^{11}$ cm$^{2}$,
where the interactions are stronger.\cite{rem:lsda}

\begin{figure}[h]
\hspace*{-.1cm}
\epsfxsize=3.6in
{\centerline{\epsffile{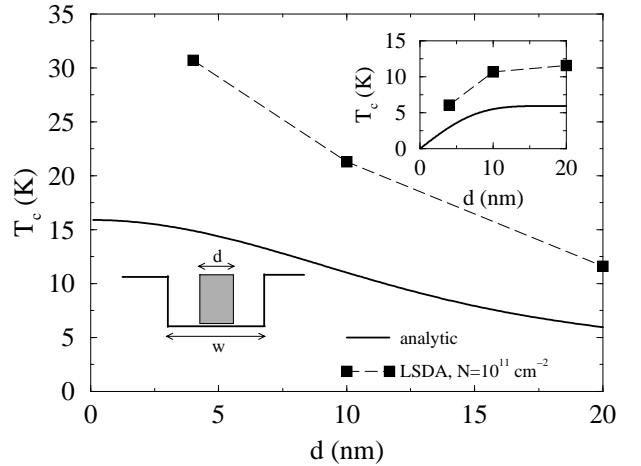}}}
\caption{Dependence of $T_c$ on the central portion $d$ of the    
quantum well ($w=20$~nm) occupied by magnetic ions.  The main graph
is for fixed 2D magnetic ion density, $c\ d=20$~nm$^{-2}$, 
while the inset is for the 
case of fixed 3D magnetic ion density, $c=1$~nm$^{-3}$.}
\label{fig:two}
\end{figure}

Before discussing the implications of
the expression (\ref{tc}) for $T_c$ we comment on its relation to the RKKY
theory\cite{dietl} of carrier induced ferromagnetism in DMS's.
As applied to bulk 3D electron systems the two approaches are
fundamentally equivalent, although there are differences in
detail which can be important.  The principle difference is that
the RKKY theory treats the free-carrier magnetic-ion interaction
perturbatively.  As a result both approaches give the same value for the 
critical temperature where the magnetization vanishes;
differences appear at lower temperatures where the electron system
is strongly or completely spin-polarized and a perturbative
treatment of its interactions with the magnetic ions fails.
Both approaches fail to account for the retarded character of
the free-carrier mediated interaction between magnetic ions 
in these low density systems, and for correlated quantum
fluctuations in magnetic
ion and free carrier subsystems which are important for some
properties.\cite{jurgenhsiuhau}  Both approaches are able to
account for interaction effects in the free carrier system, which
increase the tendency toward ferromagnetism as illustrated in
Fig.~\ref{fig:one}, and
in the density-functional approach appear in the spin-dependence
of the exchange correlation potential.  It is in inhomogeneous
cases, such as the reduced dimension situation considered here,
that the density functional approach has an advantage.  Our
approach provides a simple description of the collective behavior
of the quasi-2D ferromagnet and is able to handle geometric 
features which can be engineered to produce desired properties.
For example we show in Fig.~\ref{fig:two} a plot of the dependence
of the critical temperature on the fraction $d$ of the quantum well
occupied by magnetic ions. When the integrated 2D density of magnetic
ions ($c\ d$) is fixed, the critical temperature increases as  magnetic
impurities are transferred toward the center of the quantum well where the free-carrier
wavefunction has its maximum. For  fixed 3D density $c$ the number of magnetic
impurities decreases with decreasing $d$. This effect is stronger than the increase 
of the overlap between the carrier wavefunction and magnetic ion distribution resulting
in the decrease of $T_c$, as shown in the inset of Fig.~\ref{fig:two}. In both
cases the analytic results of Eq.~(\ref{tc}) are in qualitative agreement with
the full numerical LSDA calculations.

\begin{figure}[h]
\hspace*{-.1cm}
\epsfxsize=3.6in
{\centerline{\epsffile{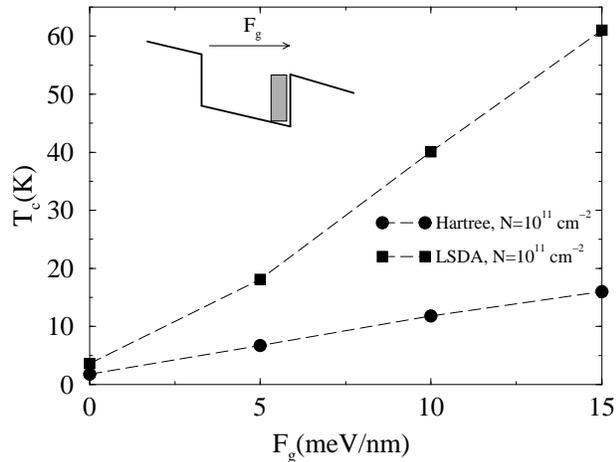}}}
\caption{Dependence of $T_c$ on bias voltage applied across  the 
$w=10$~nm wide quantum well partially occupied 
by magnetic ions over a distance of 3~nm near one edge of the quantum well.
Circles (squares) represent results of the full numerical self-consistent 
calculations without (with) the exchange-correlation potential.}
\label{fig:three}
\end{figure}

One remarkable feature of quasi-2D ferromagnetic
DMS's is the possibility of tuning $T_c$ through a wide range
in situ, by the application of a gate voltage.  In
Fig.~\ref{fig:three} we illustrate this for the case of a 
$w=10$~nm  quantum well with magnetic ions covering
only a $d=3$~nm portion near one edge.  
Even for such a narrow quantum
well, the critical temperature can be varied over an order of 
magnitude by applying a bias voltage which draws  electrons into
the magnetic ion region.

In closing we remark that for several reasons
DMS ferromagnetism mediated by
2D electron systems is likely to occur only at inaccessibly
low temperatures.  One important factor is the 
smaller value of the exchange interaction parameter in the conduction
band case.
For example for the typical n-type II-VI DMS
quantum well$^3$, (Zn$_{1-x}$Cd$_{x}$Se)$_{m-f}$(MnSe)$_f$,\cite{2dsmorchkova} 
the $s-d$ exchange $J_{sd} \sim 4\times10^{-3}$ eV\ nm$^{3}$.
Our theory then gives $T_c \sim 10$~mK for a 10~nm wide quantum well.
Equally important, however, is the weak magnetic anisotropy expected in the conduction 
band case, which will lead to soft spin-wave collective modes\cite{jkm} and 
substantial $T_c$ suppression.

This work was supported by the National Science Foundation under
grants DMR-9714055 and DGE-9902579,  and
by the Grant Agency of the Czech Republic
under grant 202/98/0085.

\begin{references}
\bibitem{caveatanisotropy} More generally there is a delicate interplay
between size-quantization controlled heavy-light
subband mixing and kinetic-exchange interaction anisotropy which
can be accounted for if the geometry of a particular sample is
accurately known.
\bibitem{rem:lsda} The LSDA form for the exchange-correlation potential would
diverge in the limit of infinitely narrow 2D system. For the 10 nm wide quantum we
expect the LSDA to overestimate the effect of interactions
among carriers and, hence, to overestimate the critical temperature.
Accurate theories of valence-band exchange and correlation effects are
an important challenge for theory.
\end{references}
\end{document}